\documentclass[prl,aps,twocolumn]{revtex4}

\usepackage{graphicx}
\usepackage{amsmath}
\usepackage{color}
\usepackage{units}
\usepackage[latin1]{inputenc}
\usepackage{verbatim}
\usepackage{float}
\usepackage[english]{babel}

\begin{document}

\title{Experimental entanglement distribution by separable states}

\author{Christina E.~Vollmer,$^1$ Daniela Schulze,$^1$ Tobias Eberle,$^1$ Vitus H\"andchen,$^1$ Jarom\'{i}r Fiur\'{a}\v{s}ek$^2$ and Roman Schnabel$^{1,*}$}

\affiliation{
$^1$Institut f\"ur Gravitationsphysik, Leibniz Universit\"at Hannover and Max-Planck-Institut f\"ur Gravitationsphysik (Albert-Einstein-Institut), Callinstr.~38, 30167~Hannover, Germany\\
$^2$Department of Optics, Palack\'y University, 17.\ listopadu 50, 77200 Olomouc,
Czech Republic\\
$^*$Corresponding author: roman.schnabel@aei.mpg.de
}

\maketitle

\textbf{The distribution of entanglement between macroscopically separated parties represents a crucial protocol for future quantum information networks \cite{Kimble08, Horodecki2009}.
Surprisingly, it has been theoretically shown that two distant systems can be entangled by sending a third mediating system that is not entangled with either of them \cite{Cubitt2003,Mista2008}. Such a possibility seems to contradict the intuition that to distribute entanglement, the transmitted system always needs to be entangled with the sender.
Here, we experimentally distribute entanglement by exchanging a subsystem and successfully prove that this subsystem is not entangled with either of the two parties.  
Our implementation relies on the preparation of a specific three-mode Gaussian state containing thermal noise that demolishes the entanglement in two of the three bipartite splittings.
After transmission of a separable mode this noise can be removed by quantum interference. 
Our work demonstrates an unexpected variant of entanglement distribution and improves the understanding necessary to engineer multipartite quantum information networks.} 

\begin{figure}
\includegraphics[width=\columnwidth]{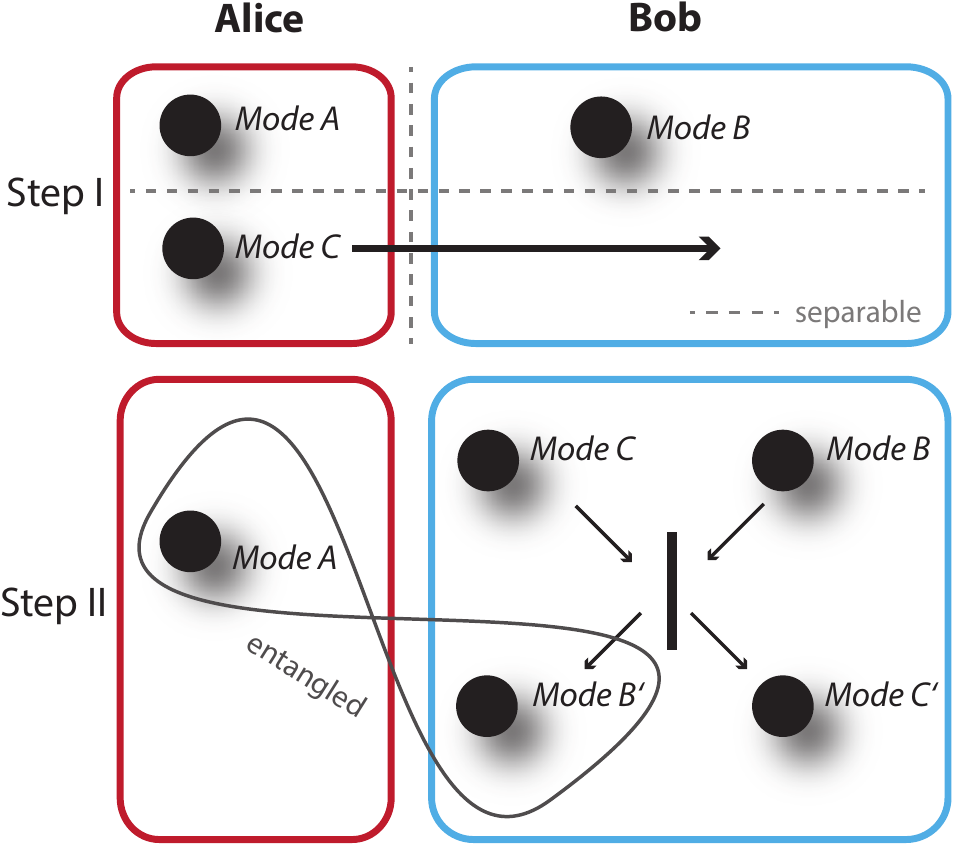}
\caption{\textbf{Principle of entanglement distribution by separable states.} In the beginning Alice possesses the two separable modes A and C. Both modes are also separable with respect to Bob's mode B. Alice sends mode C to Bob and he combines his mode B with the received mode C. Finally, Alice and Bob share an entangled system $\text{A}|\text{B}^\prime$, which can be traced back to the initial entanglement for the $\text{A}|\text{B}\text{C}$ splitting.}
\label{fig: principle}
\end{figure}

The principle of entanglement distribution by separable states is illustrated in Fig.~\ref{fig: principle}. 
In the beginning of the protocol Alice possesses two separable modes A and C, while Bob possesses the mode B, which is also separable from Alice's modes. In a first step Alice sends the ancilla mode C, which is neither entangled with mode A nor with mode B, to Bob. To obtain two-mode entanglement Bob mixes his modes B and C in the second step of the protocol. One output mode is then discarded, while the other one turns out to be entangled with A. The distributed entanglement can be used for further quantum information protocols \cite{Braunstein2005}, such as quantum teleportation \cite{Furusawa98, Bowen2003} and quantum key distribution \cite{Scarani2009}.

This remarkable and seemingly paradoxical protocol is made possible by a specific structure of quantum correlations in the underlying three-mode Gaussian state. For the protocol to work the state must be separable with respect to the $\text{B}|\text{A}\text{C}$ and $\text{C}|\text{A}\text{B}$ splittings and inseparable with respect to the $\text{A}|\text{B}\text{C}$ splitting.
According to the classification introduced in \cite{Giedke2001}, we therefore need a three-mode Gaussian entangled state belonging to Class III. 

Our protocol thus explores the rich structure of multimode entanglement, which can exhibit more complex properties and features than two-mode entanglement and which represents a valuable resource for lots of applications ranging from local realism tests \cite{Zhao2003} to one-way quantum computing \cite{Raussendorf01,Walther2005,Ukai11}.
\begin{figure}
\includegraphics[width=\columnwidth]{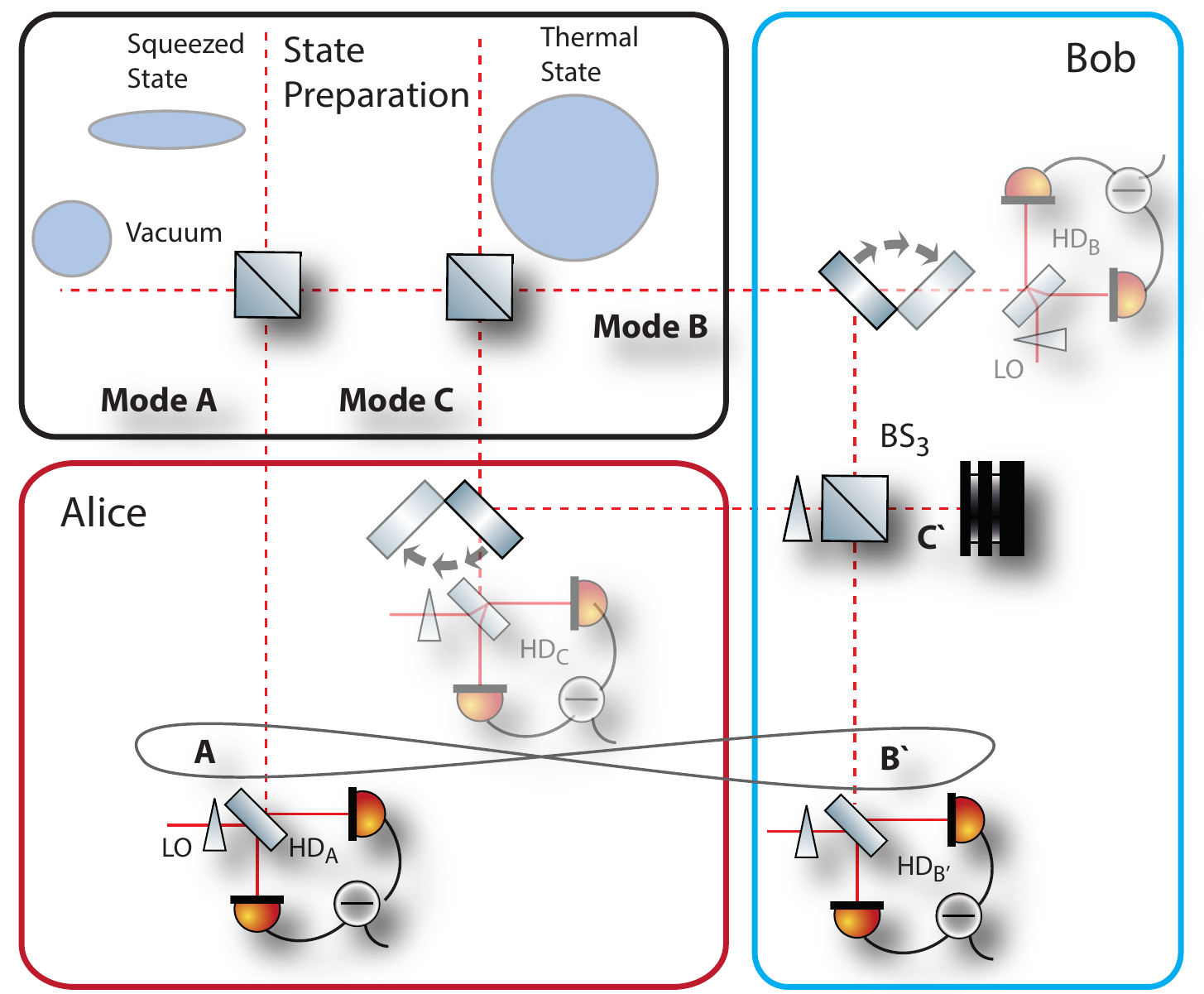}
\caption{\textbf{Experimental setup of the entanglement distribution by separable states.} The three-mode state is prepared by overlapping a squeezed state, a vacuum state and a thermal state at two balanced beam splitters.  After modes A and C had been sent to Alice and mode B had been sent to Bob, the separability properties were checked with balanced homodyne detectors $\text{HD}_\text{A}$, $\text{HD}_\text{B}$ and $\text{HD}_\text{C}$. Afterwards Alice sent the separable mode C to Bob. By overlapping the modes C and B at another balanced beam splitter $\text{BS}_3$, entanglement was established between Alice and Bob, which was verified with homodyne detectors $\text{HD}_\text{A}$ and $\text{HD}_{\text{B}^\prime}$.}
\label{fig: aufbau}
\end{figure}
\newline
\newline
\textbf{Three-mode state preparation}\\
Our experimental setup is depicted in Fig.~\ref{fig: aufbau}. The initial three-mode Gaussian state is prepared by an independent source and is distributed between Alice and Bob. The preparation starts with a squeezed state, which interferes with a vacuum state at a balanced beam splitter. 
The beam splitter output A is sent directly to Alice, while the other output is superimposed with a thermal state at a second balanced beam splitter. 
After the state preparation Alice possesses modes A and C, while Bob holds mode B. The separability properties of this three-mode state (ABC) were checked by a tomographic reconstruction of the full three-mode covariance matrix with the homodyne detectors $\text{HD}_\text{A}$, $\text{HD}_\text{B}$ and $\text{HD}_\text{C}$ and found to be separable with respect to the $\text{B}\vert \text{A}\text{C}$ and $\text{C}\vert \text{B}\text{A}$ splittings.
In the next step Alice sends the mode C to Bob, where the modes B and C interfere at a balanced beam splitter with the appropriate phase to get rid of the correlated noise. This step creates two-mode entanglement between Alice and Bob, which is verified by measuring the Duan criterion \cite{Duan2000}
\begin{equation}
\text{Var}(\hat{X}_{\text{A}}-\hat{X}_{\text{B}^\prime})+\text{Var}(\hat{P}_{\text{A}}+\hat{P}_{\text{B}^\prime})<4
\label{eq: Duan}
\end{equation}
using the homodyne detectors $\text{HD}_\text{A}$ and $\text{HD}_{\text{B}^\prime}$. Here, $\hat{X}$ and $\hat{P}$ describe the amplitude and phase quadrature operators, respectively. They are normalized to the shot noise level, i.~e. $\text{Var}(\hat{X})=\text{Var}(\hat{P})=1$ for a vacuum state.
\newline
\newline
\textbf{Requirements for the three-mode state}\\
For investigating the separability properties of the three-mode state (ABC) we apply the positive partial transposition criterion (PPT) \cite{Horodecki1996, Simon01} to the measured state. This criterion is both necessary and sufficient for bipartite splittings of Gaussian states with $N$ modes with only a single mode on one side ($1\vert N-1$)\cite{Wolf2000}. The three-mode state is separable with respect to mode $k$ if the corresponding covariance matrix of the partially transposed state $\gamma^{\text{T}(k)}$ fulfills the uncertainty relation
\begin{equation}
\gamma^{\text{T}(k)}-i\Omega\geq 0\,,
\label{eq:PPT}
\end{equation}
with $\Omega = \bigoplus_{k=1}^3 J,$ where $J= \begin{pmatrix} 0&-1\\1&0 \end{pmatrix}$. This criterion is equivalent to finding the symplectic eigenvalues of the covariance matrix of the partially transposed state. If the smallest symplectic eigenvalue $\mu_k$, in the following called PPT value, is below 1, the state is inseparable with respect to the $k|ij$ splitting. For details see supplemental material.

We named the PPT values for the different splittings after the single mode: $\text{PPT}_{\text{A}}$ for the $(\text{A}\vert \text{B}\text{C})$ splitting, $\text{PPT}_{\text{B}}$ for $(\text{B}\vert \text{A}\text{C})$, and $\text{PPT}_{\text{C}}$ for $(\text{C}\vert \text{A}\text{B})$. Our protocol thus requires $\text{PPT}_{\text{A}}<1$ (= inseparable) and $\text{PPT}_{\text{B}}, \text{PPT}_{\text{C}} > 1$ (= separable) to verify the appropriate three-mode state for distributing entanglement by separable states.

Within the experimental setup we can vary the following critical parameters: the size of the thermal state as well as the variances of the squeezed and anti-squeezed quadratures of the squeezed state. The latter two can be changed independently of each other by variation of the pump power of the squeezed light source and by variation of additional losses. 

\begin{figure}
\includegraphics[width=\columnwidth]{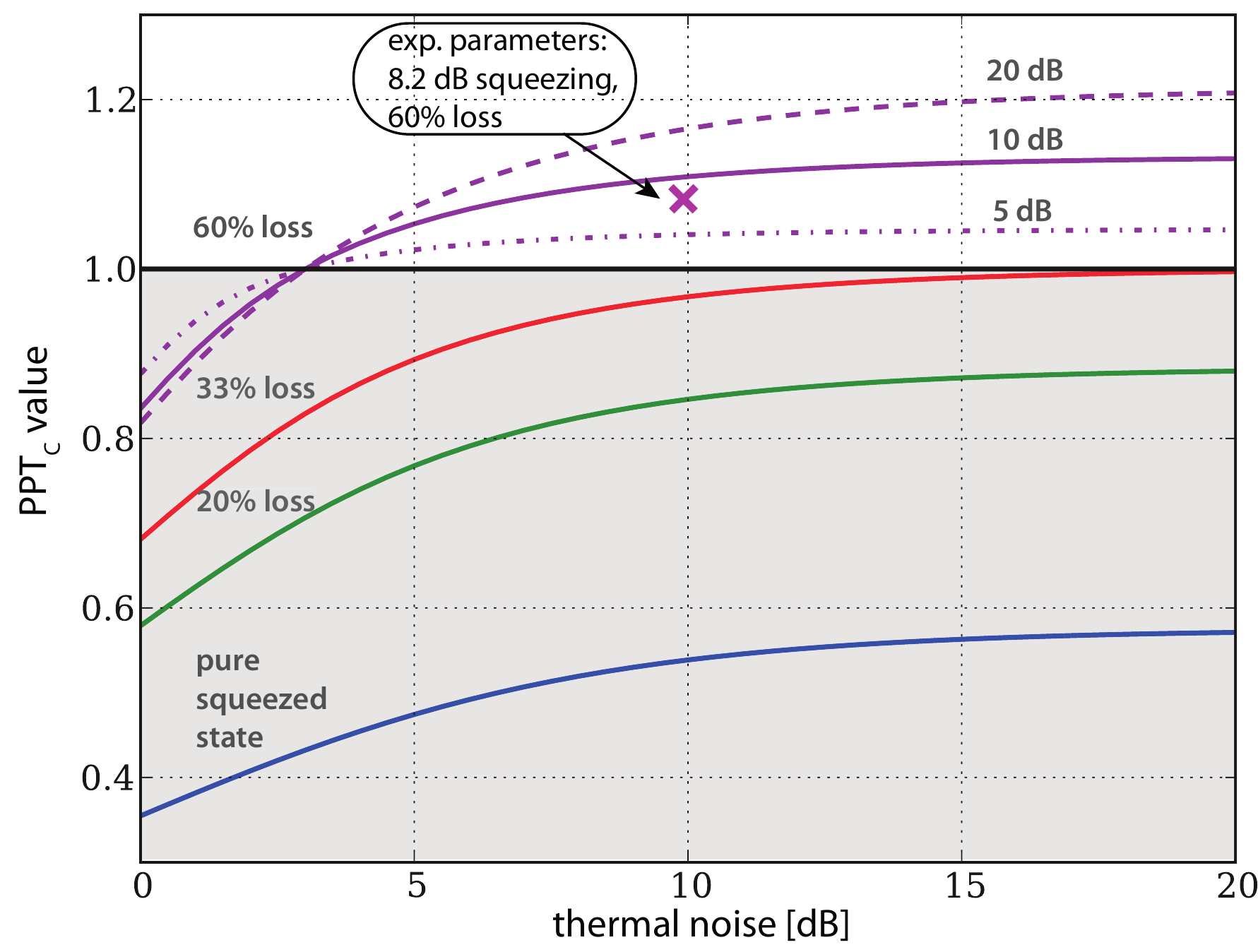}
\caption{\textbf{Separability analysis.} Theoretical simulations of the $\text{PPT}_{\text{C}}$ value with respect to the added thermal noise are depicted. The solid lines correspond to 10 dB initial squeezing with different losses. To obtain the required PPT value ($>1$) a minimal loss of 33\% is necessary. The point of intersection with the threshold of 1 is exclusively depending on the loss and not on the actual initial squeezing value. The parameters, which were used in our measurements, are marked with the magenta cross.}
\label{fig: theo1}
\end{figure}

We performed theoretical simulations to analyse the influence of the squeezed and thermal states on the separability properties of the generated three-mode state. 
Figure \ref{fig: theo1} shows the $\text{PPT}_{\text{C}}$ value of the three-mode state versus the noise power of the thermal state. Due to the symmetry of the setup $\text{PPT}_{\text{B}}$ and $\text{PPT}_{\text{C}}$ are identical. The $\text{PPT}_{\text{A}}$ value is not depicted, since this value is always below 1 when a squeezed state is used as an input state $\text{A}_\text{in}$. 

The magenta lines show the independence of the initial squeezing value on the intersection point with the separability threshold unity. Hence, the amount of classical noise necessary to demolish the entanglement only depends on the optical loss applied to the squeezed state. Nevertheless, with higher squeezing values the three-mode state is farther pushed into the separable regime with respect to the $\text{B}\vert \text{A}\text{C}$ and $\text{C}\vert \text{B}\text{A}$ splittings for a sufficiently large thermal state.  Apart from that, higher squeezing values also distribute more entanglement to the two distant parties. 

From Fig.~\ref{fig: theo1} it is also visible that the entanglement can only be demolished by classical noise if the optical loss applied to the squeezed state is larger than 33.3\%. This is exactly the border for which the bipartite entangled state, generated by the superposition of the squeezed and the vacuum state, is no longer EPR-entangled \cite{Eberle2010}. EPR-entangled states are a subclass of general entanglement, exhibiting stronger quantum correlations. Indeed the properties of our three-mode state show that these correlations are so strong that the entanglement in the bipartite splittings cannot be demolished by classical noise. 
\newline
\newline
\textbf{Experimental Results}\\ 
The 21 independent elements of the symmetric $6\times6$ three-mode covariance matrix are determined from homodyne measurements on modes A, B, and C. For each quadrature measurement we recorded $10^6$ data points. 
As input states we used a squeezed state with -1.8\,dB and 5.1\,dB noise reduction/amplification in the amplitude and phase quadrature, respectively, and an elliptical thermal state (hot squeezed state) with 9.6\,dB and 10.2\,dB noise amplification. The resulting three-mode covariance matrix $\gamma$ was measured as
\begin{equation*}\gamma= 
\begin{pmatrix}
0.76 & 0.04 & 0.12 & -0.03 & 0.19 & -0.07\\
0.04 & 2.20 & 0.05 & -0.78 & -0.10 & -0.74\\
0.12 & 0.05 & 5.70 & -0.29 & -3.92 & 1.14\\
-0.03 & -0.78 & -0.29 & 6.84 & -0.96 & -3.94\\
0.19 & -0.10 & -3.92 & -0.96 & 4.73 & 0.09\\
-0.07 & -0.74 & 1.14 & -3.94 & 0.09 & 5.92	
\end{pmatrix}.
\end{equation*} 
 
\begin{figure}
\includegraphics[width=\columnwidth]{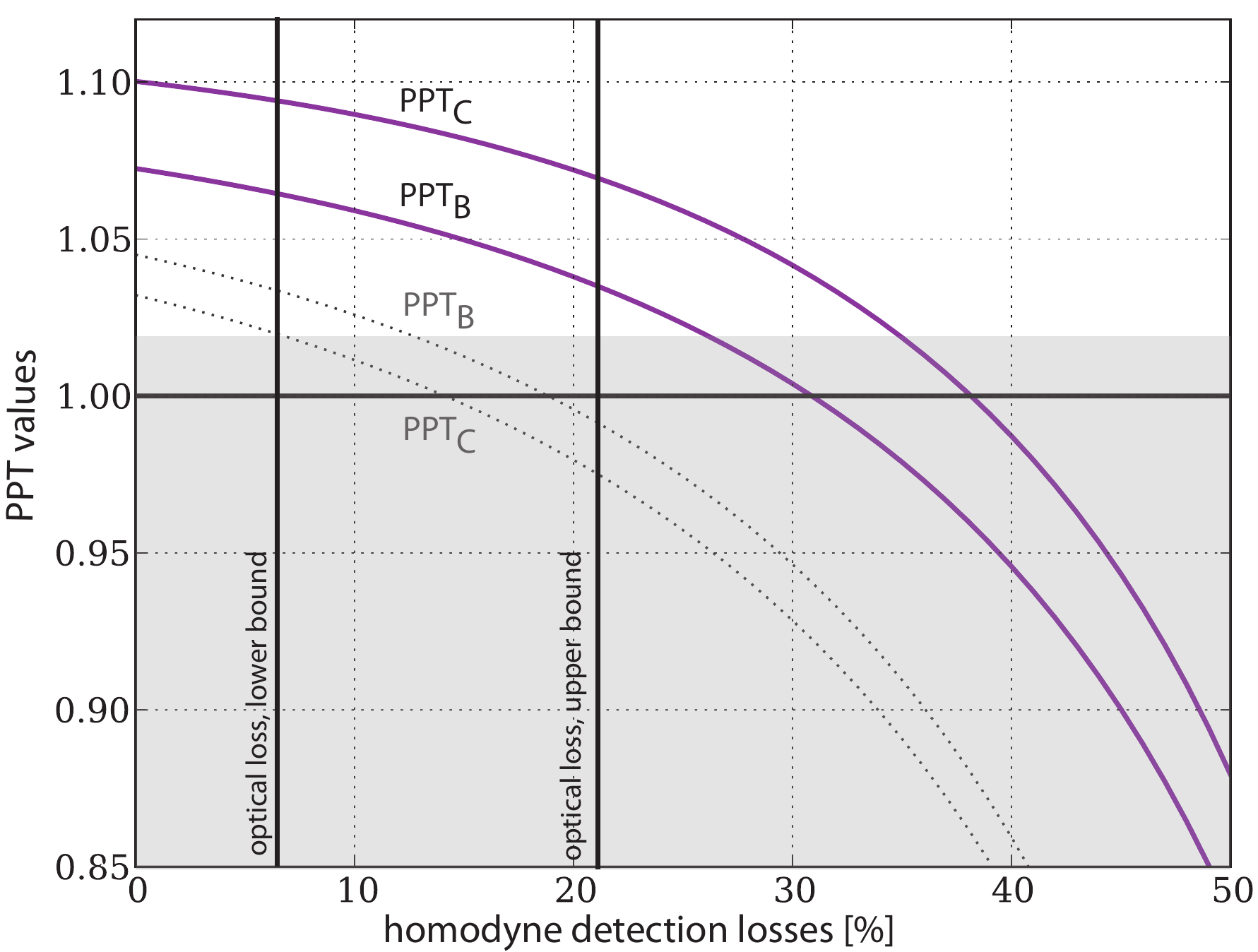}
\caption{\textbf{Measured PPT values with subtraction of detection losses.} The magenta curves show the inferred $\text{PPT}_\text{B}$ and $\text{PPT}_\text{C}$ values of the measured covariance matrix $\gamma$ for a spectrum of computationally eliminated detection losses. Based on independent measurements we estimate the actual detection loss to be greater 7\% and smaller 22\%. These losses do not push the PPT values below unity. The successful demonstration of our protocol is thus independent of the question whether detection loss should be corrected for or not.
In contrary, the grey dotted curves, which shows the PPT values for another measurement, fall below the threshold, when losses $> 15\%$ are subtracted and thus do not fulfill the three-mode state requirements. }
\label{fig: theoloss}
\end{figure}
This covariance matrix directly leads to the PPT values: $\text{PPT}_\text{A}=0.89$, $\text{PPT}_\text{B}=1.1$ and $\text{PPT}_\text{C}=1.07$. Thus, the measured state fulfilled the requirements for distributing entanglement via separable states.

Three main effects could in principle cause masking the actual presence of entanglement. Two of them can also lead to a non-Gaussian state and can thus prohibit the application of the separability criterion for Gaussian states: phase fluctuations due to imperfect phase locking between signal beams and local oscillator beams and the generation of the thermal state by random displacements of originally squeezed states, where the distribution of random displacements can be non-Gaussian. These effects are considered in detail in the appendix with the result that none of them has any non-negligible effect in the presented measurement.

The third effect is the influence of detection losses. Since we are in fact interested in the separability properties of the state \textit{before} homodyne detection, the optical loss introduced by the measurement devices has to be computationally eliminated. Indeed, the separability properties of the state can be altered by a non-perfect detection process as depicted in Fig.~\ref{fig: theoloss}.
The blue curves represent the $\text{PPT}_\text{B}$ and $\text{PPT}_\text{C}$ values of the covariance matrix $\gamma$, if optical loss within the homodyne detection is subtracted. The black vertical lines mark the regime of our estimated detection efficiency (quantum efficiency + visibility).
We estimate the quantum efficiency of the homodyne's photodiodes to be about 90\%. The visibilities of the homodyne detectors were measured before each measurement and laid in a regime of 93-98\%. For the covariance matrix $\gamma$ the detection losses thus are $11\%\pm 5\%$ for the homodyne detector $\text{HD}_\text{A}$, $17\%\pm 5\%$ for $\text{HD}_\text{B}$ and $16.6\%\pm 5\%$ for $\text{HD}_\text{C}$, which leads to an upper bound of 6\% and a lower bound of 22\% loss. 
After subtracting the losses from the three-mode covariance matrix the lower and upper bounds for the PPT values are 0.85 and 0.87 for the $\text{A}\vert \text{B}\text{C}$ splitting, 1.07 and 1.09 for the $\text{B}\vert \text{A}\text{C}$ splitting and 1.04 and 1.06 for the $\text{C}\vert \text{A}\text{B}$ splitting and thus fulfill the criteria. 
This shows the correctness of the separability properties regardless whether the detection loss is considered to be part of the detected state or not.

The grey curves in Fig.~\ref{fig: theoloss} however represent the PPT values of an inappropriate example, where the state is inseparable in all splittings, if the estimated detection losses are subtracted. Thus, this state does not allow a demonstration of entanglement distribution by separable states. 

Monte Carlo simulations show that for the $10^6$ measurements per homodyne setting the statistical error bars on the symplectic eigenvalues are of the order of 0.001. That means, that the inferred separability properties are statistically reliable even for the extreme limit of 22\% loss.

After the prepared three-mode state had been checked for its separability properties, the ancilla mode C, which was separable to the modes A and C, was sent to Bob. 
Two-mode entanglement between Alice and Bob was generated by superimposing modes C and B at the balanced beam splitter $\text{BS}_3$  with the appropriate phase, which was controlled manually. The Duan criterion resulted in 3.4 ($<4$), which proved that entanglement was successfully distributed by separable states.


In conclusion, we experimentally realized entanglement distribution by separable states. We showed that for this protocol a specific three-mode state is suitable, whose thermal noise destroys the entanglement in two bipartite splittings, which can later be restored via quantum interference. Thereby we could show, that the protocol does not work with EPR-entangled states, since with states in this class of entanglement separability cannot be generated by thermal noise.
While the entanglement distribution by separable states seems highly counterintuitive in the first place, our protocol provides an insight into the underlying physical mechanism behind this protocol. From a broader perspective our work helps to understand the possibilities and restrictions offered by mulitmode entangled quantum states and future multipartite quantum communication networks.\\

\textbf{Methods}\\
1. \textit{Squeezing Source}\\
The squeezed states which we used in our protocol were generated by parametric amplification. The squeezing source was a hemilithic optical parametric amplifier (OPA) consisting of $\chi^{(2)}$ nonlinear  7\% magnesium oxide doped lithium niobate (MgO:LiNbO3) crystal inside a standing wave cavity. The cavity was composed of the crystal's back surface with a high-reflective coating for both wavelengths and the front mirror with a reflectivity of 94\% for the squeezing field at 1064\,nm and a reflectivity of 25\% for the pump field at 532\,nm. These values correspond to a finesse of 100 for 1064\,nm and 4.3 for 532\,nm. The free spectral range was about 4\,GHz due to the cavity's length of 6.5\,mm. The cavity was stabilized by the Pound-Drever-Hall technique with a sideband frequency of 30\,MHz. Therefore, a dim control field of about 2\,mW entered the cavity from the back end and the length was controlled by a piezo electric transducer driving the end mirror of the cavity. The crystal's temperature was actively stabilized at about $60^\circ$C for the phase matching of the fundamental and the harmonic field.

2. \textit{Generation of the thermal state}\\
The thermal state, which we used as the input state $\text{C}_\text{in}$, was generated as a classical squeezed state, also called a \textit{hot squeezed state}. For this state preparation we also used an optical parametric amplifier (OPA) as described above, which reduced the noise in the phase quadrature and amplifies noise in the amplitude quadrature. An electro-optical modulator applied a random noise on the phase quadrature of the vacuum mode entering the OPA. The random displacement resulted from the output of a homodyne detector measuring shot noise. Therefore, we obtained a state with a noise distribution far above the vacuum noise in all quadratures. For details see \cite{DiGuglielmo2011}.\\

\textbf{Acknowledgements}\\
This research was supported by the FP7 project QESSENCE (Grant agreement number 248095) and the Centre for Quantum Engineering and Space-Time Research (QUEST). We thank the IMPRS on Gravitational Wave Astronomy for support. V.H. acknowledges support from HALOSTAR.
J.F. acknowledges support from the Czech Science Foundation under project P205/12/0694. 
\newline
\section{Appendix}

\subsection{Covariance matrix}
Let $X_j$ and $P_j$ denote the amplitude and phase quadratures of mode $j$ that satisfy the canonical commutation relations $[X_j,P_k]=2i\delta_{jk}$. Here 
$\delta_{jk}$ denotes the Kronecker delta and the quadrature operators are normalized such that the variance of vacuum state quadratures is equal to $1$. When dealing with $N$
modes, it is convenient to collect the quadratures into a vector $Q=(X_1,P_1,\cdots,X_j,P_j,\cdots, X_N,P_N)$. The elements of the $N$-mode covariance matrix are defined as follows \cite{Braunstein2005,WB12},
\begin{equation}
\gamma_{jk}=\frac{1}{2}\langle \Delta Q_j \Delta Q_k+\Delta Q_k \Delta Q_j \rangle,
\end{equation}
where $\Delta Q_j=Q_j-\langle Q_j\rangle$ and $\langle \rangle$ denotes averaging over a quantum state. The covariance matrix of a physical state is symmetric and positive definite
and satisfies the generalized Heisenberg uncertainty relation \cite{Braunstein2005}
\begin{equation}
\gamma -iJ_N \geq 0,
\end{equation}
where 
\[
J_N=\bigoplus_{j=1}^N \left(
\begin{array}{cc}
0 & 1 \\
-1 & 0 
\end{array}
\right)
\]
is the $N$-mode symplectic form. Symplectic eigenvalues of $\gamma$ are defined as the positive roots of polynomial 
\begin{equation}
|\gamma- i\mu J_N|=0
\end{equation}
where $|A|$ denotes the determinant of matrix $A$.
The symplectic eigenvalues of a covariance  matrix of physical state satisfy the inequality $\mu \geq 1$.  

The entanglement and separability of Gaussian states may be conveniently analyzed with the help of the partial transposition criterion \cite{Simon01,Wolf2000}. 
At the level of quadrature operators, the partial transposition 
with respect to mode $j$ corresponds to the change of sign of the phase quadrature, $P_k \rightarrow -P_k$. The covariance matrix of a state partially transposed with respect to mode $j$
thus reads,
\begin{equation}
\gamma^{T(k)}=T_k\gamma T_k^T,
\end{equation}
where $T_k$ is a diagonal matrix with all diagonal elements equal to $1$ except for $T_{2k,2k}=-1$. Consider a bipartite splitting of an $N$-mode Gaussian state with covariance matrix $\gamma$
such that one party holds mode $k$ and the other party possesses all the remaining $N-1$ modes. As shown by Werner and Wolf \cite{Wolf2000}, the partial transposition criterion is both 
necessary and sufficient for this case. The Gaussian state is entangled with respect to the considered bipartite splitting if and only if the lowest symplectic eigenvalue $\mu_k$ of $\gamma^{(T_k)}$ satisfies $\mu_k<1$ and otherwise it is separable.

\subsection{Verification of separability}

In our experiment, we prepare a specific three-mode state and we want to demonstrate its separability with respect to the bipartite splittings
$B|AC$ and $C|AB$. Since the state is generated by mixing single-mode squeezed states on beam splitters, it seems natural to assume that the state is Gaussian and apply 
the necessary and sufficient separability criteria for Gaussian states. 
However, the entanglement of the state may be potentially undetectable in the Gaussian setting if certain experimental imperfections make the state slightly non-Gaussian.

\begin{figure}[!b!]
\centerline{\includegraphics[width=0.99\linewidth]{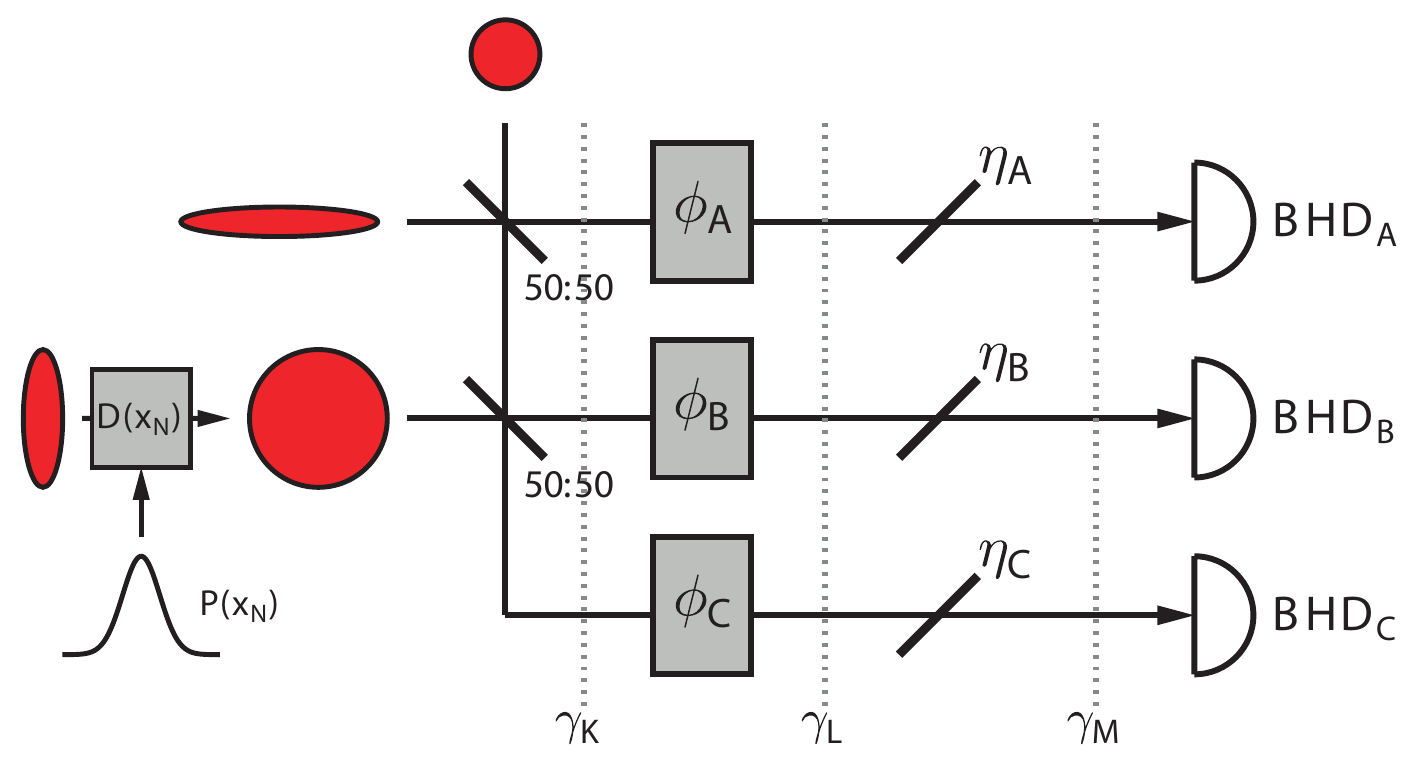}}
\caption{Equivalent scheme of the experimental setup including effects that can potentially mask entanglement of the resulting three-mode state. The state preparation begins by splitting a single-mode
squeezed vacuum on a balanced beam splitter. One mode is then mixed with a noisy quasi-thermal state produced from a squeezed vacuum state by random displacements of the initially squeezed quadrature.
After mixing on beam splitters, each mode can be subject to phase fluctuations and losses. The purely lossy channel with transmittance  $\eta_j$ followed by perfect homodyne detector 
models a realistic homodyne detector with efficiency $\eta_j$ and the phase fluctuations account for imperfect phase stability between signal and local-oscillator beams in the homodyne detector. 
Note that the phase shift operation commutes with a lossy channel hence we can consider the ordering depicted in the figure without any loss of generality.}
\end{figure}

Here, we identify the main effects that may potentially influence the assessment of the entanglement properties of the considered state 
and we demonstrate that the separability properties of the experimentally generated state remain unchanged when these effects are taken into account and compensated for.
As illustrated in Fig. S.1, the three main effects that may mask the entanglement of the state  consist of inefficient homodyne detection, phase fluctuations due to imperfect phase locking between 
signal beams and local oscillator beams, and generation of hot squeezing by random displacements of originally squeezed quadrature where the distribution of random displacements may 
slightly deviate from Gaussian distribution. In what follows, we will consider each of these effects in detail.

\subsubsection{Inefficient homodyne detection}

In the experiment, the three-mode state is probed with three balanced homodyne detectors and the covariance matrix $\gamma$ of the state is determined from 
the homodyne data. The homodyne detectors are characterized by their efficiency $\eta_j$ that is limited by the quantum efficiency of the photodiodes, the mode-matching of signal 
and local oscillator beams and the losses of the passive optical elements in the paths of the signal beams. Inefficient homodyne detection is equivalent to a combination of a transmission through
a purely lossy channel with transmittance $\eta$ followed by ideal homodyne detection. The  covariance matrix $\gamma_L$ before transmission through lossy channels 
is then related to the reconstructed covariance matrix $\gamma_M$ according to
\begin{equation}
\gamma_M=S_\eta \gamma_L S_{\eta}^T+G_\eta,
\label{losschannel}
\end{equation}
where 
\begin{equation}
S_\eta=\left(
\begin{array}{cccccc}
\sqrt{\eta_A} & 0 & 0 & 0 & 0 & 0  \\
0 & \sqrt{\eta_A} & 0 & 0 & 0 & 0  \\
0 & 0 & \sqrt{\eta_B} & 0 & 0 & 0  \\
0 & 0 & 0 & \sqrt{\eta_B} & 0 & 0  \\
0 & 0 & 0 & 0 & \sqrt{\eta_C} & 0  \\
0 & 0 & 0 & 0 & 0 & \sqrt{\eta_C}  
\end{array}
\right)
\end{equation}
and $G_\eta=I-S_\eta S_\eta^T$ where $I$ denotes the identity matrix. 
The  covariance matrix $\gamma_L$ can be determined by inverting the linear relation (\ref{losschannel}),
\begin{equation}
\gamma_L=S_\eta^{-1} (\gamma_M-G_\eta){S_\eta^{T}}^{-1}.
\label{gammaLinversion}
\end{equation}
It may happen that the state with covariance matrix $\gamma_L$ is actually entangled with respect to one or both of the splittings $B|AC$, $C|AB$ although losses induce separability of $\gamma_M$ 
with respect to both these splittings. To convincingly verify that this is not the case in our experiment, we have to take conservative estimates of $\eta_j$ which means that we should
consider the lowest possible values of $\eta_j$ consistent with our characterization of homodyne detectors and the statistical errors.  For the reconstructed covariance matrix 
\begin{equation}
\gamma_M =\left(
\begin{array}{rrrrrr}
0.76 & 0.04 & 0.12 & -0.03 & 0.19 & -0.07 \\
0.04 & 2.20 & 0.05 & -0.78 & -0.10 & -0.74 \\
0.12 & 0.05 & 5.70 & -0.29 & -3.92 & 1.14 \\
-0.03 & -0.78 & -0.29 & 6.84 & -0.96 & -3.94 \\
0.19 & -0.10 & -3.92 & -0.96 & 4.73 & 0.09 \\
-0.07 & -0.74 & 1.14 & -3.94 & 0.09 &  5.92 
\end{array}
\right)
\end{equation}
the conservative lower bounds on homodyne detection efficiencies read
\begin{equation}
\eta_A=0.839, \qquad  \eta_B=0.780, \qquad \eta_C=0.784.
\end{equation} 
On inserting the above explicit values into formula (\ref{gammaLinversion}) we obtain
\begin{equation}
\gamma_L=\left(
\begin{array}{rrrrrr}
%
 0.71 &   0.05  &  0.15 &  -0.04  &  0.23 &  -0.09 \\
 0.05 &   2.43  &  0.06 &  -0.96  & -0.12 &  -0.91 \\
 0.15 &   0.06  &  7.03 &  -0.37  & -5.01 &   1.46 \\
-0.04 &  -0.96  & -0.37 &   8.49  & -1.23 &  -5.04 \\
 0.23 &  -0.12  & -5.01 &  -1.23  &  5.76 &   0.11 \\
-0.09 &  -0.91  &  1.46 &  -5.04  &  0.11 &  7.28  
   \end{array}
\right).
\end{equation}
The separability of a Gaussian state with covariance matrix $\gamma_L$ can be checked by calculating minimum symplectic eigenvalue $\mu_j$ 
of a covariance matrix corresponding to the density matrix of a partially transposed state $\rho^{T_j}$, $j=A,B,C$. For the matrix $\gamma_L$ we obtain
\begin{equation}
\mu_A=0.85, \qquad \mu_B=1.07, \qquad \mu_C=1.04.
\end{equation}
We have $\mu_B>1$ and $\mu_C>1$ which confirms the separability with respect to the splittings $B|AC$ and $C|AB$. We should also check that the 
loss compensation does not lead to an unphysical covariance matrix. 
An explicit calculation confirms that $\gamma_L$ satisfies the generalized Heisenberg inequality $\gamma_L+iJ \geq 0$ and the lowest symplectic eigenvalue of $\gamma_L$ 
reads  $\mu_0= 1.11>1$,
hence $\gamma_L$ is a physically allowed covariance matrix. 

If the true homodyne detection efficiencies $\tilde{\eta}_{j}$ would be larger than the lower bounds $\eta_j$, then the true covariance matrix would be obtained by sending $\gamma_L$ through local lossy channels with transmittances $\eta_j/\tilde{\eta}_j$. Since local losses cannot generate any entanglement, the $B|AC$ and $C|AB$ separability of the state is preserved for all actual homodyne detection efficiencies higher than the lower bounds $\eta_j$.

\subsubsection{Phase fluctuations}
Besides losses and inefficient detection, the homodyne detection can be also influenced by random phase fluctuations between signal and local-ocillator beams. 
A single-mode phase shift by angle $\phi$ couples the amplitude and phase quadratures $X$ and $P$,
\begin{equation}
X'=X\cos\phi+P\sin\phi, \qquad P'=P\cos\phi-X\sin\phi.
\end{equation}
We can conveniently describe this canonical transformation by a $2\times 2$ orthogonal matrix
\begin{equation}
R(\phi)= \left( \begin{array}{cc} \cos\phi & \sin\phi \\ -\sin\phi & \cos\phi \end{array}\right),
\end{equation}
multiplying the column quadrature vector $(X,P)^T$.

We shall consider independent random phase shift $\phi_j$ for each mode, $j=A,B,C$. The three-mode phase shift operation is thus described by a block diagonal $6\times 6$ matrix
\begin{equation}
S(\boldsymbol\phi) = R(\phi_A)\oplus R(\phi_B)\oplus R(\phi_C),
\end{equation}
where $\boldsymbol\phi=(\phi_A,\phi_C,\phi_D)$. Let $\gamma_K$ denote the covariance matrix  in the absence of phase fluctuations. 
The phase shift $S(\boldsymbol\phi)$ transforms the covariance matrix $\gamma_K$ as follows,
\begin{equation}
\gamma_K^\prime=S(\boldsymbol\phi) \gamma_K S^T(\boldsymbol\phi).
\end{equation}
Since the mean values of all quadratures of the studied state vanish,  the observed covariance matrix $\gamma_L$ can be determined by averaging the phase shifted covariance matrix $\gamma_K^\prime$ over the random phase shifts $\phi$,
\begin{equation}
\gamma_L= \left\langle S(\boldsymbol\phi) \, \gamma_K \, S^T(\boldsymbol{\phi}) \right\rangle_{\boldsymbol\phi}.
\label{gammaKaverage}
\end{equation}
As there is no reason for correlated phase fluctuations in our experiment, we shall assume that the random phase shifts are independent 
and mutually uncorrelated. To be specific, we shall assume that each phase shift obeys Gaussian distribution with zero mean and variance $\sigma^2$. The relevant averages then read
\begin{equation}
\langle\cos\phi_j\rangle=e^{-\sigma^2/2}, \qquad \langle \sin \phi_j\rangle = 0,
\label{average1}
\end{equation}
\begin{equation}
\langle\cos^2\phi_j\rangle=\frac{1+e^{-2\sigma^2}}{2}, \qquad \langle \sin^2 \phi_j\rangle = \frac{1-e^{-2\sigma^2}}{2}.
\label{average2}
\end{equation}
Note that in the experimentally relevant limit of small phase fluctuations the exact shape of the probability distribution becomes irrelevant and the above formulas may be considered
as universally valid in the limit $\sigma \ll 1$.
With the help of the expressions (\ref{average1}), (\ref{average2}), and (\ref{gammaKaverage}) we obtain 
\begin{equation}
\gamma_L = e^{-\sigma^2}\gamma+\frac{(1-e^{-\sigma^2})^2}{2} \mathcal{D}(\gamma_{K})
+\frac{(1-e^{-2\sigma^2})}{2}J \mathcal{D}(\gamma_{K})J^T.
\label{KLrelation}
\end{equation}
Here $J$ denotes the three-mode symplectic form and $\mathcal{D}(\gamma_{K})$ denotes a covariance matrix that is obtained from $\gamma_K$ 
by setting all covariances between quadratures of different modes equal to zero. 
Hence $\mathcal{D}(\gamma_{K})$ has a block diagonal structure with three $2\times 2$ matrices on the main diagonal. The formula (\ref{KLrelation}) can be inverted and after some algebra one finds
\begin{equation}
\gamma_K = e^{\sigma^2}\gamma_L+\frac{(e^{\sigma^2}-1)^2}{2} \mathcal{D}(\gamma_{L}) +\frac{1-e^{2\sigma^2}}{2} J \mathcal{D}(\gamma_{L})J^T.
\end{equation}
This expression allows us to undo the effect of phase fluctuations and determine the covariance matrix of a state before it was affected by the phase noise, c.f. Fig.~S.1.

\begin{figure}[!t!]
\centerline{\includegraphics[width=0.99\linewidth]{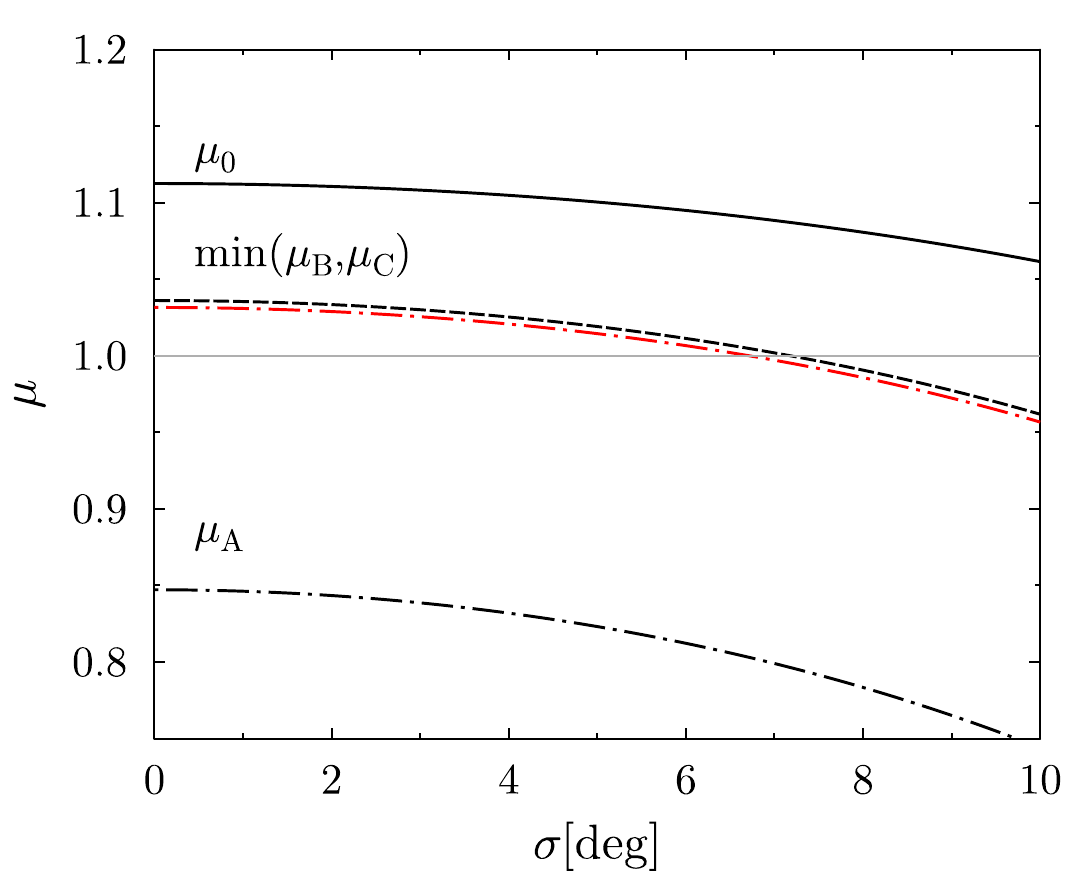}}
\caption{The lowest symplectic eigenvalue $\mu_0$ of covariance matrix $\gamma_K$ (solid black line), symplectic eigenvalue $\mu_A$ indicating entanglement in the $A|BC$ splitting 
(black dot-dashed line) and the parameter $\min(\mu_B,\mu_C)$ (black dashed line). We can see that the state remains $B|AC$ and $C|AB$ separable if the strength of random phase fluctuations
is below $\sigma_{\mathrm{th}}\approx 7^{\circ}$. The red dot-dashed line represents plot of $\min(\mu_B,\mu_C)$ for covariance matrix $\gamma_0$ with removed non-Gaussian fraction of noise
added by random modulation in hot squeezing generation. See text for details.}
\end{figure}

If the state with covariance matrix $\gamma_K$ is $B|AC$ and $C|AB$ separable, then the local phase fluctuations preserve this property because local single-mode operations
cannot generate any entanglement from a separable state. 
  In Fig. S.2 we plot the symplectic eigenvalues characterizing the entanglement properties of $\gamma_K$ 
as a function of the strength of phase fluctuations $\sigma$. We can see that a Gaussian state with covariance matrix $\gamma_K$ remains $B|AC$ and $C|AB$ separable unless the 
phase fluctuations exceed the threshold value of $\sigma_{\mathrm{th}}\approx 7^{\circ}$. 

Note that the phase-diffused state with covariance matrix $\gamma_L$ can be slightly non-Gaussian 
due to phase fluctuations. Therefore, the covariance-matrix based separability criteria valid for Gaussian states cannot be directly applied to this state 
(these criteria become only a necessary condition for separability but not a sufficient one for general non-Gaussian states). Instead, we have shown above that 
we can undo the effect of phase fluctuations on the reconstructed covariance matrix and determine the covariance matrix of a seed Gaussian state before the phase fluctuations. 
The separability properties of the phase diffused state can be then rigorously inferred by checking the separability of this seed Gaussian state.

\subsubsection{Hot-squeezing generation}
Finally, let us consider the procedure for hot-squeezing generation by random displacements of the squeezed quadrature, see Fig.~S.1. 
In the experiment, a great effort is made to achieve a Gaussian modulation
and preserve the Gaussian nature of the state. Nevertheless, slight deviations from perfectly Gaussian modulations may occur in practice and our goal here is to assess their impact.
In the Heisenberg picture, the random modulation can be described by an addition of a noisy component $x_N$ to the squeezed amplitude quadrature of mode C, 
\begin{equation}
x_C \rightarrow x_C+x_N.
\end{equation}
The random variable $x_N$ has zero mean and let $\sigma_N^2$ denote its variance.
We shall assume that the added noise $x_N$ contains both Gaussian and non-Gaussian part,
\begin{equation}
x_N=\sqrt{p_G}\,x_{N,G}+\sqrt{1-p_G}\,\tilde{x}_N.
\label{xN}
\end{equation}
Here $p_G$ denotes the fraction of Gaussian modulated noise, $x_{N,G}$ is Gaussian random variable with zero mean variance $\sigma_N^2$ and $\tilde{x}_{N,G}$ 
can exhibit  an arbitrary distribution with zero mean and variance $\sigma_N^2$.

The overall state generated by the noise addition becomes a statistical mixture of Gaussian states obtained by first adding the Gaussian part of the noise $\sqrt{p_G}x_{N,G}$ and then 
randomly coherently  displacing $x_C$ by $\sqrt{1-p_G}\,\tilde{x}_N$. 
Note that the coherent displacements do not influence the entanglement properties of the state and can be set to zero by local unitary displacement operations.
Therefore, if a three-mode Gaussian state obtained by considering only the Gaussian part of noisy modulation in Eq. (\ref{xN}) is $B|AC$ and $C|AB$ separable, then also the total state
including the full noisy modulation $x_N$  is $B|AC$ and $C|AB$ separable.

Consider now the experimentally determined covariance matrix $\gamma_K$ that already includes compensations for the inefficient homodyne detection and phase fluctuations.
Since this matrix is by definition symmetric, it can be diagonalized by a suitable orthogonal transformation $O$,
\begin{equation}
\gamma_K=O \gamma_{D}O^T,
\end{equation}
where $\gamma_D$ is a diagonal matrix with eigenvalues of $\gamma_K$ on the main diagonal. Given our method of preparation of the three-mode state by mixing uncorrelated single-mode states
on balanced beam splitters, the six eigenvalues of $\gamma_K$ should correspond to the variances of squeezed and anti-squeezed quadratures of the three input states, respectively. 
For instance, for $\sigma=0.05$ the six eigenvalues explicitly read
\[
\begin{array}{lllll}
\lambda_1=0.57, &~~~ & \lambda_3=1.18, & ~~~& \lambda_5= 11.55, \\
\lambda_2=3.88, & & \lambda_4=1.28, & & \lambda_6=13.23.
   \end{array}
\]
We can see that the eigenvalues have the expected structure. Eigenvalues $\lambda_1$ and $\lambda_2$ correspond to the variances of squeezed and anti-squeezed quadratures of squeezed vacuum state in mode B. The variances $\lambda_3$ and $\lambda_4$ correspond to the vacuum state input in mode B. Ideally, we should have $\lambda_3=\lambda_4=1$ for vacuum state but due to various experimental imperfections the eigenvalues are slightly larger. Finally, the last two eigenvalues $\lambda_5$ and $\lambda_6$ correspond to the highly noisy hot-squeezed pseudothermal state
injected in mode C.

We calculate the covariance matrix $\gamma_0$ of the effective Gaussian state obtained by considering only the Gaussian part $x_{N,G}$ of the random modulation $x_N$ by 
replacing $\lambda_{j}$ with $p_G\lambda_{j}$ where $j=5$ or $j=6$. Since it is difficult to reliably identify which of the two variances $\lambda_5$, $\lambda_6$ 
corresponds to the randomly modulated amplitude, we conservatively rescale both these eigenvalues. 
Let $\tilde{\gamma}_D$ denote the covariance matrix with  $\lambda_5$ and $\lambda_6$ replaced with $p_G\lambda_{5}$ and $p_G\lambda_{6}$, respectively. Then
\begin{equation}
\gamma_0=O \tilde{\gamma}_{D} O^T.
\end{equation}
We calculate the minimum symplectic eigenvalues $\mu_B$ and $\mu_C$ that characterize the separability properties of Gaussian state with covariance matrix $\gamma_0$ 
with respect to $B|AC$ and $C|AB$ splittings, respectively. The parameter $\min(\mu_B,\mu_C)$ is plotted as a red dot-dashed line in Fig. S.2. In this example we set $p_G=0.75$, i.e. we assume $25\%$
of non-Gaussian modulation. We can see that our experiment is very robust against this non-Gaussian modulation and $25\%$ of non-Gaussianity only very slightly reduces the maximum tolerable
phase fluctuations.
We estimate that in the actual experiment the non-Gaussian fraction of the random modulation is much smaller than $25\%$.
The dependences of the symplectic eigenvalues $\mu_0$ and $\mu_A$  of $\gamma_0$ on the strength of phase noise $\sigma$ practically coincide with the results obtained for  $\gamma_K$ 
and therefore we do not plot them as separate curves in Fig. S.2.
This coincidence is in full agreement with theoretical predictions because both $\mu_A$ and $\mu_0$ are chiefly determined by the properties of the squeezed vacuum state in mode A 
and are not sensitive to change of fluctuations in input mode C provided that the noise in mode C remains high above the shot noise level. 

\subsection{Statistical errors}
The three-mode covariance matrix $\gamma_M$ is reconstructed from the quadrature measurements of the homodyne detectors. The quadrature operator measured by each homodyne detector 
$X_j\cos\theta_j +P_j\sin\theta_j$ is determined by the relative phase $\theta_j$  between local oscillator and signal beam.
Amplitude quadrature $X_j$ is measured for $\theta_j=0$ and phase quadrature $P_j$ is recorded for $\theta_j=\pi/2$. In order to probe the correlations between amplitude and phase quadratures of a given mode
we also use the setting $\theta_j=\pi/4$ which corresponds to measurement of $X_j+P_j$. In the experiment, we perform measurements for six different settings of the three measured quadratures
as indicated in Table S.I. These six settings are sufficient to determine all matrix elements of $\gamma_M$. The variances and covariances of amplitude quadratures $X_j$ 
are determined in the first step and this procedure is repeated also for the phase quadratures $P_j$ in the second step. In the third (fourth, fifth) step we determine covariances 
of phase quadrature $P_A$ ($P_B$, $P_C$) with the amplitude quadratures of the other two modes.
Finally, the final sixth step yields the intramodal correlations $\langle \Delta X_j\Delta P_j+\Delta P_j\Delta X_j\rangle$. 
A total number of $10^6$ data was recorded for each measurement setting.

\begin{table}[t!]
\caption{Homodyne detector settings for complete reconstruction of the three-mode covariance matrix.}
\begin{ruledtabular}
\begin{tabular}{cccc}
 setting \#  &  BHD$_A$ & BHD$_B$ & BHD$_C$  \\
\hline 1 & $X_A$  & $X_B$ & $X_C$ \\
2 & $P_A$  & $P_B$ & $P_C$  \\
3 & $P_A$  & $X_B$ & $X_C$ \\
4 & $X_A$  & $P_B$ & $X_C$ \\
5 & $X_A$  & $X_B$ & $P_C$ \\
6 & $X_A+P_A$  & $X_B+P_B$ & $X_C+P_C$  
\end{tabular}
\end{ruledtabular}
\end{table}

To determine the statistical errors of the symplectic eigenvalues $\mu_j$ calculated from the estimated covariance matrices, we have performed a Monte Carlo simulation of the whole experiment assuming Gaussian statistics
of the measured quadratures. For each run of the Monte Carlo simulation we have reconstructed the covariance matrix and calculated the symplectic eigenvalues. This procedure
was repeated $1000$ times which provided a statistical ensemble for each estimate $\mu_j$. The mean values and statistical errors determined with the use of these ensembles read
\begin{equation}
\begin{array}{c}
\mu_A=0.849\pm 0.001, \\  
 \mu_B=1.069 \pm 0.001, \\
  \mu_C=1.036 \pm 0.001.
\end{array}
\end{equation}
This confirms that the statistical errors are very small compared to the deviations of $\mu_j$ from $1$ hence the observed separability properties are statistically significant.



\end{document}